\documentclass[12pt]{iopart}

\usepackage{graphicx}
\usepackage{subfig}
\usepackage{comment}
\usepackage[countmax]{subfloat}

\begin{document}

\title[Enhancement of superconductivity in BiS$_2$-based superconductors]{Enhancement of superconductivity near the pressure-induced semiconductor-metal transition in BiS$_2$-based superconductors \textit{Ln}O$_{0.5}$F$_{0.5}$BiS$_2$ (\textit{Ln} = La, Ce, Pr, Nd)}

\author{C T Wolowiec$^{1,2}$, B D White$^{1,2}$, I Jeon$^{1,2,3}$, D Yazici$^{1,2}$, \\K Huang$^{1,2,3}$ and M B Maple$^{1,2,3}$}
\address{$^1$ Department of Physics, University of California, San Diego, La Jolla, California 92093, USA}
\address{$^2$ Center for Advanced Nanoscience, University of California, San Diego, La Jolla, California 92093, USA} 
\address{$^3$ Materials Science and Engineering Program, University of California, San Diego, La Jolla, California 92093, USA} 
\ead{mbmaple@.ucsd}

\begin{abstract}
Measurements of electrical resistivity were performed between 3 and 300 K at various pressures up to 2.8 GPa on the BiS$_2$-based superconductors \textit{Ln}O$_{0.5}$F$_{0.5}$BiS$_2$ (\textit{Ln} = Pr, Nd). At lower pressures, PrO$_{0.5}$F$_{0.5}$BiS$_{2}$ and NdO$_{0.5}$F$_{0.5}$BiS$_{2}$ exhibit superconductivity with critical temperatures \textit{$T_\mathrm{c}$} of 3.5 and 3.9 K, respectively.  As pressure is increased, both compounds undergo a transition at a pressure $P_\mathrm{t}$ from a low \textit{$T_\mathrm{c}$} superconducting phase to a high \textit{$T_\mathrm{c}$} superconducting phase in which \textit{$T_\mathrm{c}$} reaches maximum values of 7.6 and 6.4 K for PrO$_{0.5}$F$_{0.5}$BiS$_{2}$ and NdO$_{0.5}$F$_{0.5}$BiS$_{2}$, respectively. The pressure-induced transition is characterized by a rapid increase in \textit{$T_\mathrm{c}$} within a small range in pressure of $\sim$0.3 GPa for both compounds.  In the normal state of PrO$_{0.5}$F$_{0.5}$BiS$_{2}$, the transition pressure $P_\mathrm{t}$  correlates with the pressure where the suppression of semiconducting behaviour saturates. In the normal state of NdO$_{0.5}$F$_{0.5}$BiS$_{2}$, $P_\mathrm{t}$ is coincident with a semiconductor-metal transition. This behaviour is similar to the results recently reported for the \textit{Ln}O$_{0.5}$F $_{0.5}$BiS$_2$ (\textit{Ln} = La, Ce) compounds. We observe that $P_\mathrm{t}$ and the size of the jump in \textit{$T_\mathrm{c}$} between the two superconducting phases both scale with the lanthanide element in \textit{Ln}O$_{0.5}$F$_{0.5}$BiS$_2$ (\textit{Ln} = La, Ce, Pr, Nd). 
 \end{abstract}

  % pressure effects on crystal structure 61.50.Ks; transport in superconductors; 74.25.F-; pressure effects on superconducting transition temperature 74.62.Fj;  non-cuprate materials multinary compounds, 74.70.Dd; Materials effects on transition temperature (superconductivity), 74.62.Bf


\submitto{\JPCM}
\maketitle

\section{Introduction}
\label{sec:introduction}
\indent The recent discovery of the BiS$_2$-based superconductor Bi$_4$O$_4$S$_3$ by Mizuguchi \textit{et al.} \cite{Mizuguchi1,Singh} with a superconducting critical temperature \textit{$T^{\mathrm{onset}}_\mathrm{c}$} of 8.6 K has generated much interest in a new family of BiS$_2$-based superconductors. The members of the class of novel  BiS$_2$-based superconductors have layered crystal structures that consist of superconducting BiS$_2$ layers separated by blocking layers which act as charge reservoirs that dope the BiS$_2$ layers with charge carriers.\cite{Mizuguchi1} Experimental efforts on the BiS$_2$-based materials have focused on increasing the charge carrier concentration via chemical substitution within the blocking layer \cite{Li,Jha,Jha1,Deguchi,Awana,Demura,Mizuguchi2,Xing,Yazici,Yazici2} as well as through a reduction of the unit cell volume via the application of an external pressure.\cite{Kotegawa,Wolowiec,Selvan,Selvan2} \\
\indent Recent studies of the BiS$_2$-based compounds involving chemical substitution within the blocking layers have lead to the discovery of the related superconductors \textit{Ln}O$_{0.5}$F$_{0.5}$BiS$_{2}$ (\textit{Ln} = La, Ce, Pr, Nd, Yb).\cite{Jha,Demura,Mizuguchi2,Xing,Yazici,Selvan} The compound LaO$_{0.5}$F$_{0.5}$BiS$_{2}$, synthesized under high pressure, has been reported by Mizuguchi \textit{et al.} to have a \textit{$T_\mathrm{c}$} of 10.6 K which exceeds that of Bi$_4$O$_4$S$_3$.\cite{Mizuguchi2} In addition, the compounds with \textit{Ln} = Ce, Pr, Nd, and Yb exhibit \textit{$T^{\mathrm{onset}}_\mathrm{c}$} values of $\sim$3.0 , 5.6 , 4.3 , and 5.3 K, respectively.\cite{Jha,Demura,Xing,Yazici} More recent work demonstrates that chemical substitution of the tetravalent ions Th$^{+4}$, Hf$^{+4}$, Zr$^{+4}$ and Ti$^{+4}$ for trivalent La$^{+3}$ in LaOBiS$_2$ increases the charge-carrier density and induces superconductivity.\cite{Yazici2}\\
\indent Measurements of the pressure dependence of the electrical resistivity $\rho$ and the superconducting critical temperature \textit{$T_\mathrm{c}$} have also recently been reported for several of these new compounds.\cite{Kotegawa,Wolowiec,Selvan,Selvan2} In this paper, we report the temperature dependence of the electrical resistivity $\rho$ from 3 to 300 K for the \textit{Ln}O$_{0.5}$F$_{0.5}$BiS$_{2}$ (\textit{Ln} = Pr, Nd) compounds under applied quasi-hydrostatic pressure up to $\sim$2.8 GPa. Both compounds exhibit the same qualitative evolution of \textit{$T_\mathrm{c}$} in which they undergo a pressure-induced transition at $P_\mathrm{t}$ from a low \textit{$T_\mathrm{c}$} superconducting phase to a high \textit{$T_\mathrm{c}$} superconducting phase. This transition region is characterized by a rapid increase of \textit{$T_\mathrm{c}$} in a narrow range of pressure $\sim$0.3 GPa.  In the high \textit{$T_\mathrm{c}$} phase at $\sim$2.5 GPa, we observed a maximum \textit{$T_\mathrm{c}$} of 7.6 K for PrO$_{0.5}$F$_{0.5}$BiS$_2$ and 6.4 K for NdO$_{0.5}$F$_{0.5}$BiS$_2$. In the normal state of both materials, there is a significant suppression of semiconducting behaviour with pressure which is continuous up to the pressure $P_\mathrm{t}$. A rapid increase of the charge carrier density is inferred from both the suppression of the semiconducting behaviour and the rapid increase of \textit{$T_\mathrm{c}$} in this region.\cite{Igawa}\\
\indent The pressure dependence of $\rho$ and the evolution of \textit{$T_\mathrm{c}$} reported in this article for the \textit{Ln}O$_{0.5}$F$_{0.5}$BiS$_{2}$ (\textit{Ln} = Pr, Nd) compounds are qualitatively similar to the behaviour we recently reported for the \textit{Ln}O$_{0.5}$F$_{0.5}$BiS$_{2}$ (\textit{Ln} = La, Ce) compounds.\cite{Wolowiec} For the four BiS$_2$-based layered superconductors, \textit{Ln}O$_{0.5}$F$_{0.5}$BiS$_2$ (\textit{Ln} = La, Ce, Pr, Nd), the transition pressure $P_\mathrm{t}$ and the size of the ``jump'' in \textit{$T_\mathrm{c}$} between the two superconducting phases both scale with the lanthanide element \textit{Ln}; specifically, as the atomic number of \textit{Ln} increases, $P_\mathrm{t}$ increases while the ``jump''  in \textit{$T_\mathrm{c}$} decreases.
\section{Experimental section}
\label{sec:experiment}
Polycrystalline samples of \textit{Ln}O$_{1-x}$F$_{x}$BiS$_{2}$ (\textit{Ln} = Pr, Nd) with $x$ = 0.5 were prepared by solid-state reaction using powders of Pr$_{2}$O$_{3}$ (99.9\%), PrF$_{3}$ (99.9\%), Pr$_{2}$S$_{3}$ (99.9\%), and Bi$_{2}$S$_{3}$ (99.9\%) for PrO$_{1- x}$F$_{x}$BiS$_{2}$, and powders of Nd$_{2}$O$_{3}$ (99.9\%), NdF$_{3}$ (99.9\%), Nd$_{2}$S$_{3}$ (99.9\%), and Bi$_{2}$S$_{3}$ (99.9\%) for NdO$_{1- x}$F$_{x}$BiS$_{2}$. Bi$_{2}$S$_{3}$ precursor powder was prepared in an evacuated quartz tube by reacting Bi (99.99\%) and S (99.9\%) at 500$^{\circ}$C for 10 hours.  The  \textit{Ln}$_{2}$S$_{3}$ (\textit{Ln} = Pr, Nd) precursor powders were  prepared in an evacuated quartz tube by reacting chunks of Pr and Nd with S grains at 800$^{\circ}$C for 10 hours. The starting materials with nominal composition \textit{Ln}O$_{0.5}$F$_{0.5}$BiS$_{2}$ (\textit{Ln} = Pr, Nd) were  weighed, thoroughly mixed, pressed into pellets, sealed in evacuated quartz tubes, and annealed at 800$^{\circ}$C for 48 hours. The products were ground, mixed for homogenization, pressed into pellets, and annealed again in evacuated quartz tubes at 800$^{\circ}$C for 48 hours. This last step was repeated again to promote phase homogeneity. X-ray powder diffraction measurements (not shown) were made using an X-ray diffractometer with a Cu K$_{\alpha}$ source to assess phase purity and to determine the lattice parameters of the \textit{Ln}O$_{0.5}$F$_{0.5}$BiS$_{2}$ (\textit{Ln} = Pr, Nd) compounds. The main diffraction peaks for the two samples can be well indexed to a tetragonal structure with space group \textit{P4}/\textit{nmm} conforming to the CeOBiS$_2$ structure. The lattice parameters for PrO$_{0.5}$F$_{0.5}$BiS$_{2}$  were determined to be \textit{a} = \textit{b} = 4.0192 \AA\ and \textit{c} = 13.4238 \AA,\ while for NdO$_{0.5}$F$_{0.5}$BiS$_{2}$ the lattice parameters are \textit{a} = \textit{b} = 4.0102 \AA\  and \textit{c} = 13.4468 \AA.\cite{Yazici}\\
\indent Measurements of $\rho(T)$ under applied pressure were performed up to  $\sim$2.8 GPa in a clamped piston cylinder pressure cell between $\sim$3 K and 300 K in a pumped $^4$He dewar. A 1:1 by volume mixture of $n$-pentane and isoamyl alcohol was used to provide a quasi-hydrostatic pressure transmitting medium.  Annealed Pt leads were affixed to gold-sputtered contact surfaces on each sample with silver epoxy in a standard four-wire configuration. The pressure dependence of \textit{$T_\mathrm{c}$} for high purity Sn (99.999\%) was measured inductively and used as a manometer for the experiments. The pressure was determined by calibrating our \textit{$T_\mathrm{c}$} data for Sn against data used in \cite{Smith69}. The width of the superconducting transition of the Sn manometer was used as a measure of the error in pressure, which was found to be on the order $\Delta P$ $\sim$$\pm$ 0.05 GPa.\\ 
\section{Results}
\label{sec:results}
%begin{comment}
\begin{figure}[h!] 
\begin{center}
\subfloat
{\label{Figure1a}
\includegraphics[scale=0.38, trim= 2.7cm 1.2cm 3.2cm 2cm, clip=true]{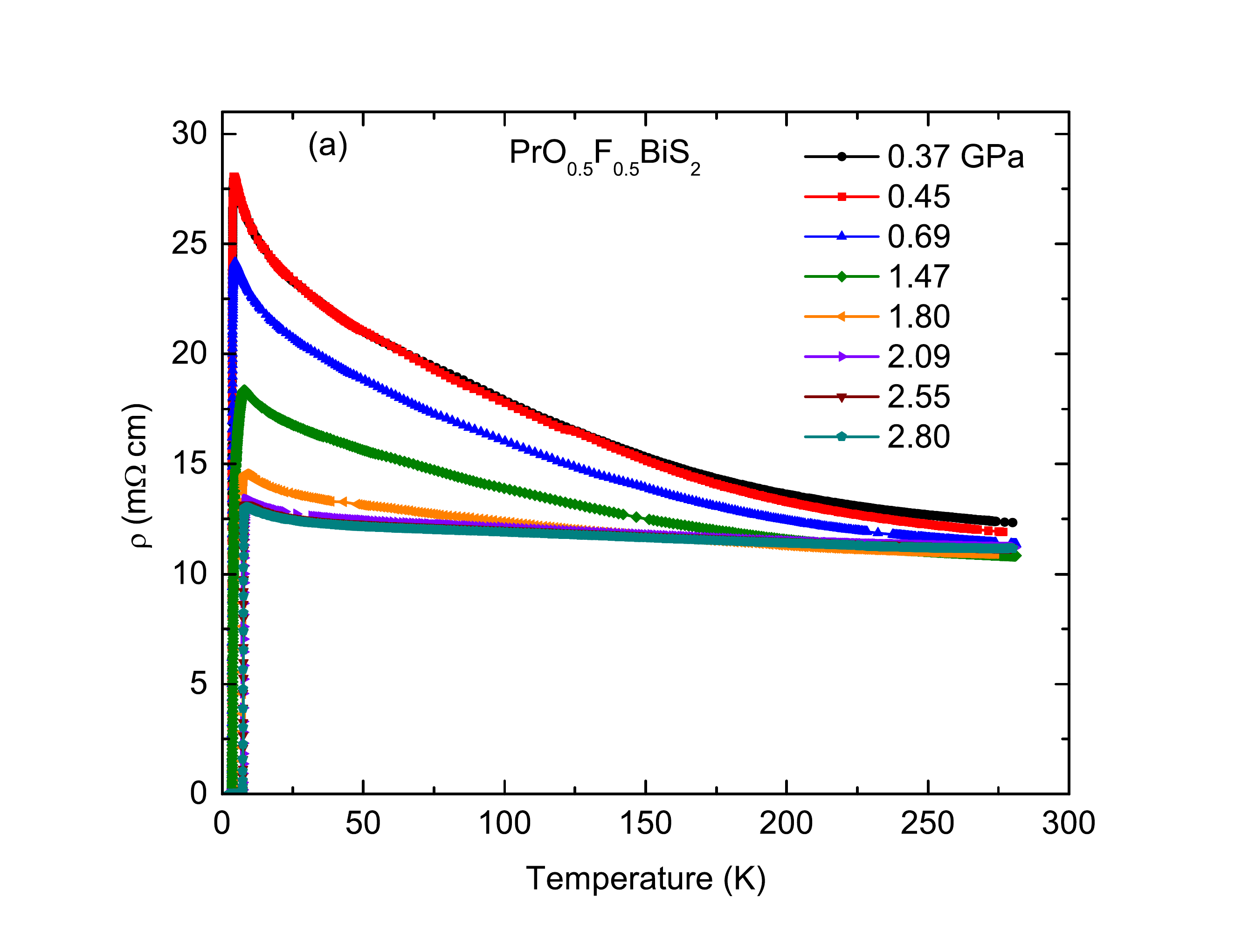}}
\subfloat
{\label{Figure1b}
\includegraphics[scale=0.38, trim= 2.8cm 1.3cm 0cm 2cm, clip=true]{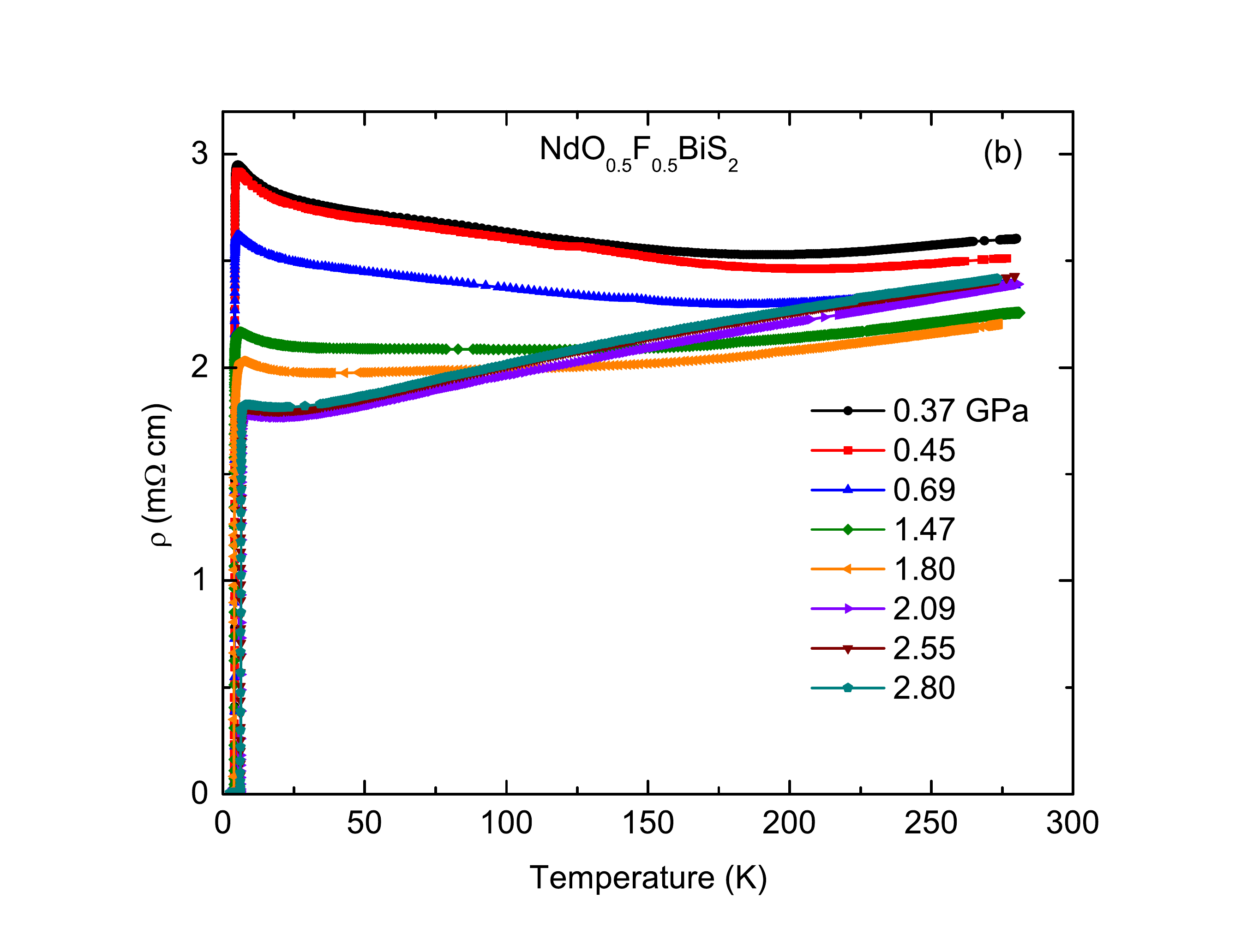}}\\
\end{center}
\subfloat
{\label{Figure1c}
\includegraphics[scale=0.38, trim= 2.7cm 1.2cm 3.2cm 2cm, clip=true]{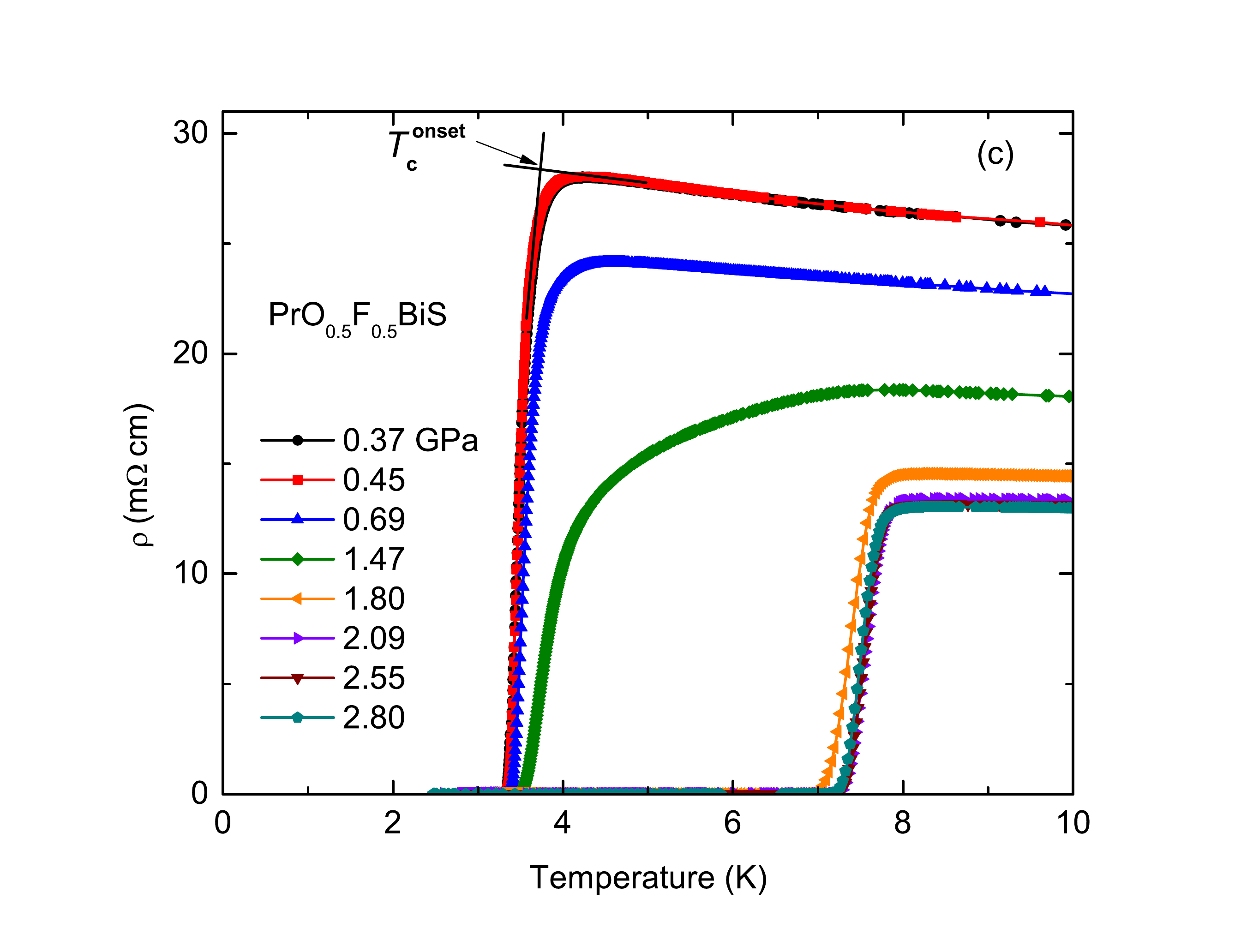}}
\subfloat
{\label{Figure1d}
\includegraphics[scale=0.38, trim= 2.8cm 1.2cm 0cm 2cm, clip=true]{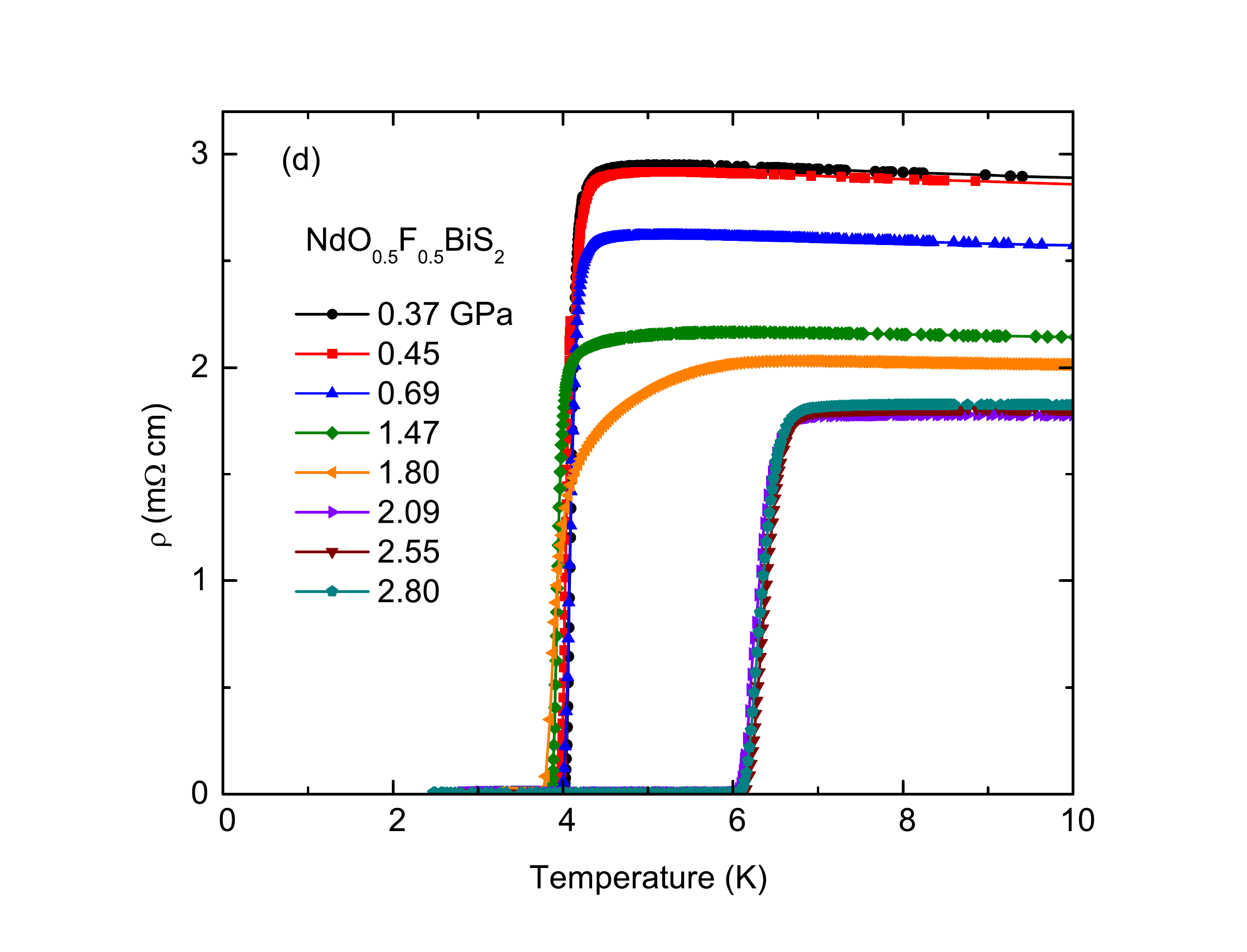}}
\centering
  \caption{\label{Resistivity} (colour online) (a),(b) Temperature dependence of the electrical resistivity $\rho$ at various pressures for (a) PrO$_{0.5}$F$_{0.5}$BiS$_{2}$ and (b) NdO$_{0.5}$F$_{0.5}$BiS$_{2}$. At lower pressures, both compounds exhibit semiconducting behaviour which is suppressed with increasing pressure. NdO$_{0.5}$F$_{0.5}$BiS$_{2}$ becomes completely metallic at $\sim$2 GPa (d$\rho$/d$T > $ 0). (c),(d) Resistive superconducting transition curves for (c) PrO$_{0.5}$F$_{0.5}$BiS$_{2}$ and (d) NdO$_{0.5}$F$_{0.5}$BiS$_{2}$ at various pressures. In PrO$_{0.5}$F$_{0.5}$BiS$_{2}$, \textit{$T_\mathrm{c}$} increases from 3.5 K  to a maximum of 7.6 K while in NdO$_{0.5}$F$_{0.5}$BiS$_{2}$,  \textit{$T_\mathrm{c}$} increases from 3.9 K  to a maximum of 6.4 K.}
\end{figure}
%end{comment}
\indent Plots of the temperature dependence of the electrical resistivity $\rho$ below 300 K for PrO$_{0.5}$F$_{0.5}$BiS$_{2}$ and NdO$_{0.5}$F$_{0.5}$BiS$_{2}$ at various pressures up to 2.8 GPa are shown in Figure~\ref{Resistivity} (a) and (b), respectively. Both compounds exhibit semiconducting behaviour at low pressure (indicated by a negative temperature coefficient of resistivity (d$\rho$/d$T < $ 0)). The semiconducting behaviour is strongly suppressed at lower pressures. As pressure is increased, the electrical resistivity $\rho$ of PrO$_{0.5}$F$_{0.5}$BiS$_{2}$ becomes weakly temperature dependent above $\sim$1.5 GPa, but remains semiconducting (d$\rho$/d$T < $ 0). In contrast, the NdO$_{0.5}$F$_{0.5}$BiS$_{2}$ sample becomes metallic at $\sim$2 GPa (indicated by a positive temperature coefficient of resistivity (d$\rho$/d$T > $ 0) in Figure~\ref{Resistivity}(b)).\\
\indent Superconducting transitions for PrO$_{0.5}$F$_{0.5}$BiS$_{2}$ and NdO$_{0.5}$F$_{0.5}$BiS$_{2}$ are displayed in Figure~\ref{Resistivity}(c) and (d), respectively. At lower pressures up to $\sim$1 GPa, the superconducting transitions in PrO$_{0.5}$F$_{0.5}$BiS$_{2}$ are grouped near 3.5 K. As pressure is increased, there is a slight broadening of the width of the superconducting transition $\Delta$\textit{$T_\mathrm{c}$} at $\sim$1.5 GPa which is immediately followed by a dramatic increase in \textit{$T_\mathrm{c}$} from $\sim$3.9 to 7.4 K in the narrow range $\sim$1.5 - 1.8 GPa. Above 1.8 GPa, \textit{$T_\mathrm{c}$} passes through a maximum of 7.6 K at $\sim$2.5 GPa and then gradually decreases with increasing pressure. 
\begin{figure}[h!]
\centering
\includegraphics[scale=0.38, trim= 2.cm 1.2cm 0cm 2cm, clip=true]{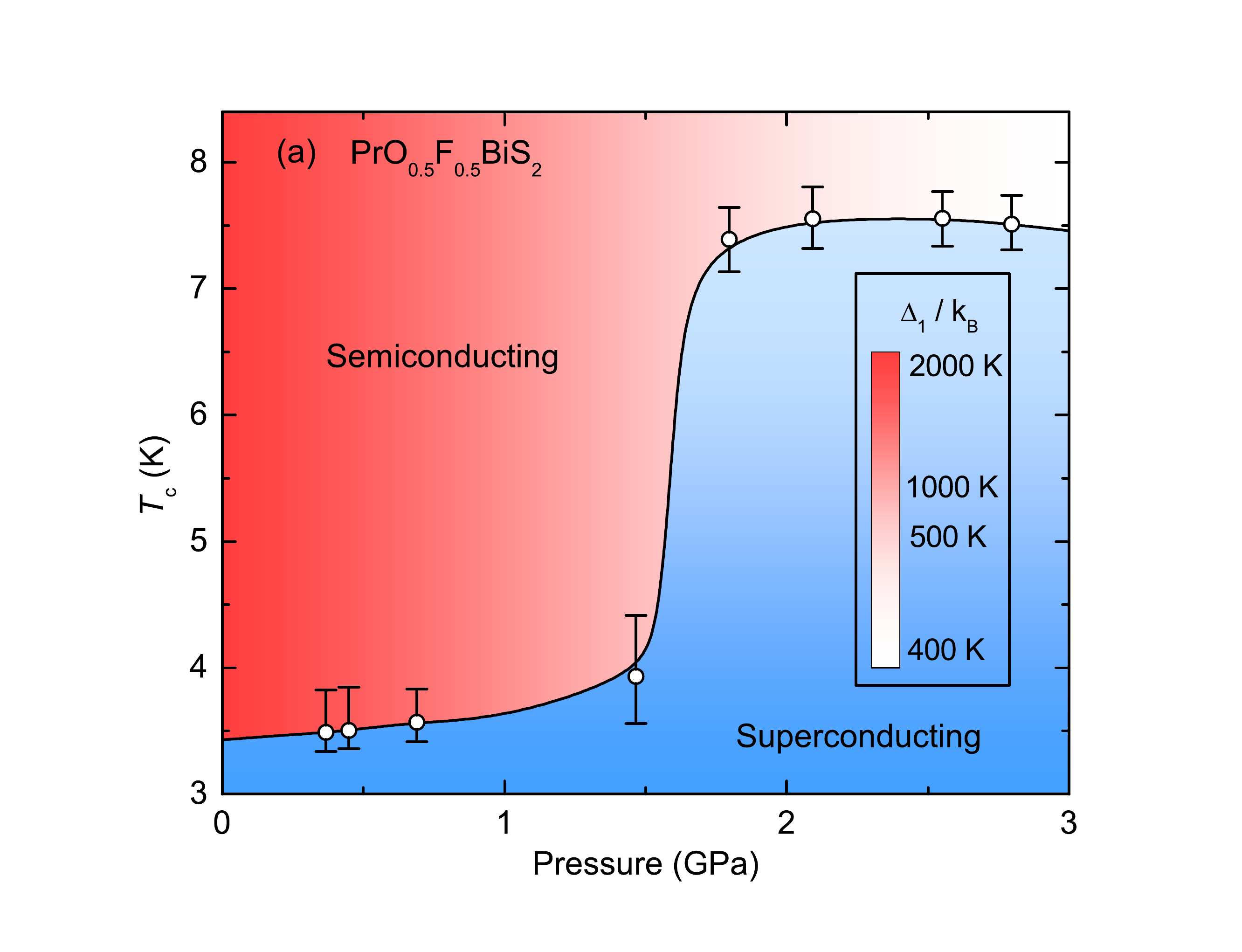}\\
\includegraphics[scale=0.38, trim= 2cm 1.2cm 0cm 2cm, clip=true]{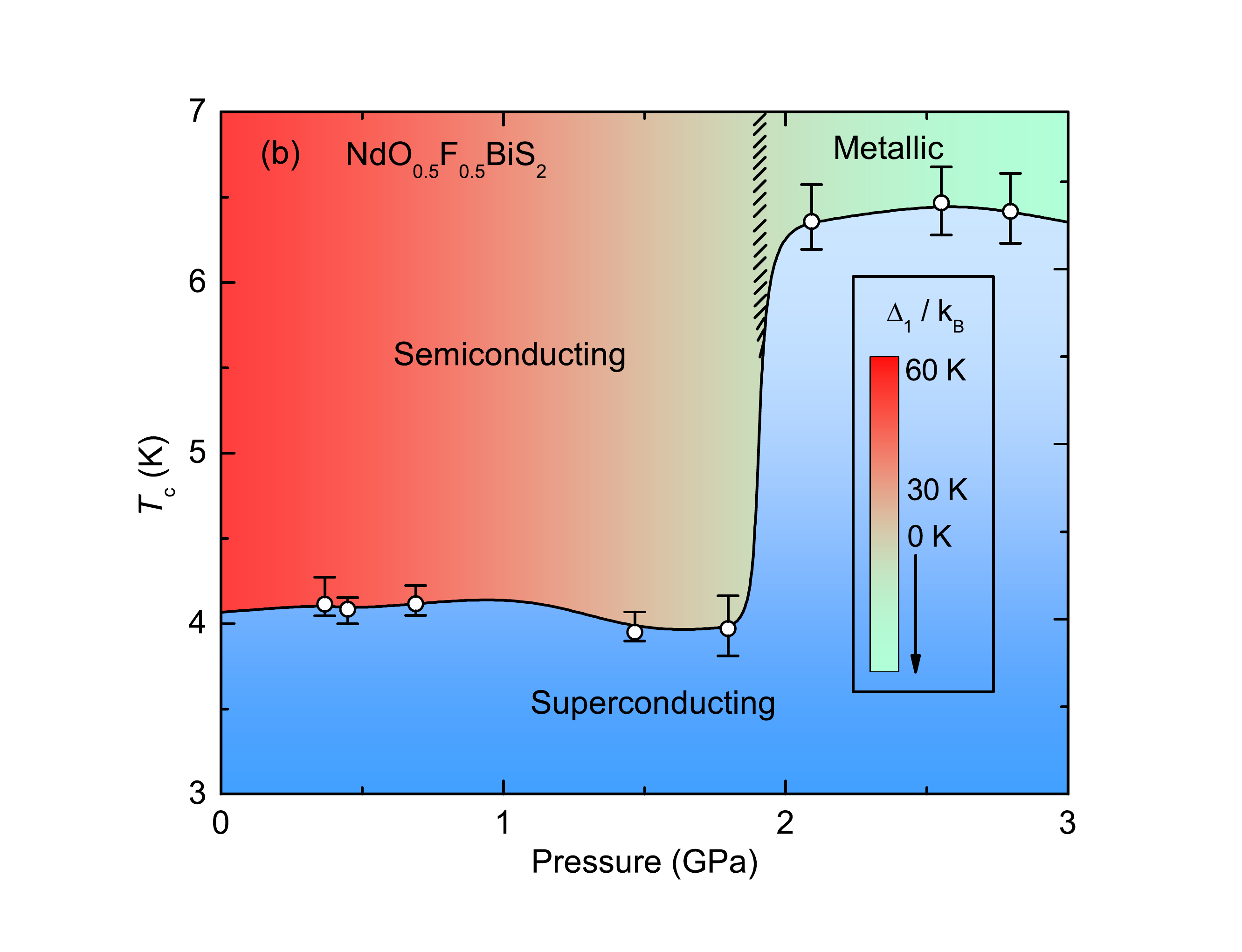}
\caption{\label{Tc Combined Spline}  (colour online) Temperature-pressure phase diagrams for (a) PrO$_{0.5}$F$_{0.5}$BiS$_{2}$ and (b) NdO$_{0.5}$F$_{0.5}$BiS$_{2}$ under pressure. Vertical bar lengths represent the transition width $\Delta$\textit{$T_\mathrm{c}$} and vertical bar caps represent \textit{$T_\mathrm{c}^{\mathrm{onset}}$} (upper) and \textit{$T_\mathrm{0}$} (lower). The colour in the semiconducting region represents the energy gap $\Delta_1$. Values for $\Delta_1$ are indicated in the false colour legend. (b) The green region to the right of the crosshatching corresponds to the metallization of NdO$_{0.5}$F$_{0.5}$BiS$_{2}$ ($\Delta_1$ = 0). The solid black curves are guides to the eye.}
\end{figure}
We observed similar behaviour in the NdO$_{0.5}$F$_{0.5}$BiS$_{2}$ compound. Sharp superconducting transitions near 4.0 K were observed at low pressures up to $\sim$1.5 GPa. The width $\Delta$\textit{$T_\mathrm{c}$} of the superconducting transition then appears to broaden near $\sim$1.8 GPa. In the small range $\sim$1.8 - 2.1 GPa, there is a sizable increase in \textit{$T_\mathrm{c}$} from $\sim$3.9 to 6.3 K.  In NdO$_{0.5}$F$_{0.5}$BiS$_{2}$, \textit{$T_\mathrm{c}$} passes through a maximum of 6.4 K at $\sim$2.5 GPa and then gradually decreases at higher pressures up to $\sim$ 2.8 GPa, similar to the behaviour observed for PrO$_{0.5}$F$_{0.5}$BiS$_{2}$.\\ 
\indent The temperature-pressure phase diagrams for the PrO$_{0.5}$F$_{0.5}$BiS$_{2}$ and NdO$_{0.5}$F$_{0.5}$BiS$_{2}$ compounds are displayed in Figure~\ref{Tc Combined Spline}(a) and (b), respectively. In the superconducting state, both compounds exhibit a low \textit{$T_\mathrm{c}$} phase which is characterized by a gradual increase in \textit{$T_\mathrm{c}$} with pressure. In the PrO$_{0.5}$F$_{0.5}$BiS$_{2}$ sample, \textit{$T_\mathrm{c}$} increases monotonically from 3.5  to 3.9 K at pressures up to $\sim$1.5 GPa (d$\textit{$T_\mathrm{c}$}$/d$P$ = 0.40 K GPa$^{-1}$). In the NdO$_{0.5}$F$_{0.5}$BiS$_{2}$ sample, there was a non-monotonic decrease in \textit{$T_\mathrm{c}$} from 4.1 to  3.9 K at pressures up to $\sim$1.8 GPa.  As pressure is increased, both compounds exhibit a rapid increase in \textit{$T_\mathrm{c}$} within a narrow range $\sim$0.3 GPa. For the PrO$_{0.5}$F$_{0.5}$BiS$_{2}$ compound, \textit{$T_\mathrm{c}$} increases dramatically from 3.9 to 7.4 K as pressure is increased from $\sim$1.5 to 1.8 GPa (d$\textit{$T_\mathrm{c}$}$/d$P$ = 11.7 K GPa$^{-1}$). 
%\begin{comment}
\begin{figure}[h!]
\centering
\includegraphics[scale=0.38, trim= 2cm 1.2cm 0cm 2cm, clip=true]{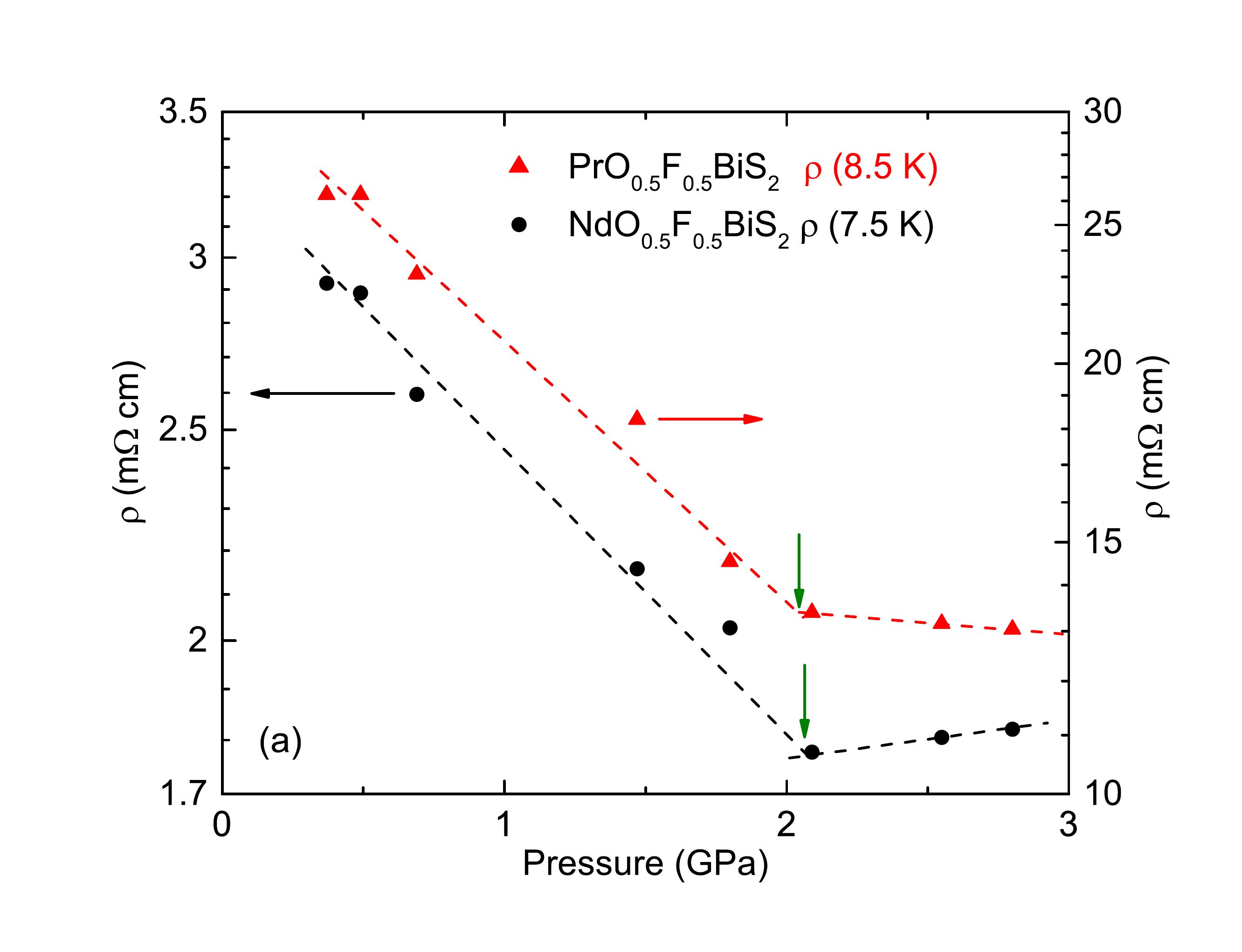}\\
\includegraphics[scale=0.38, trim= 2cm 1.2cm 0cm 2cm, clip=true]{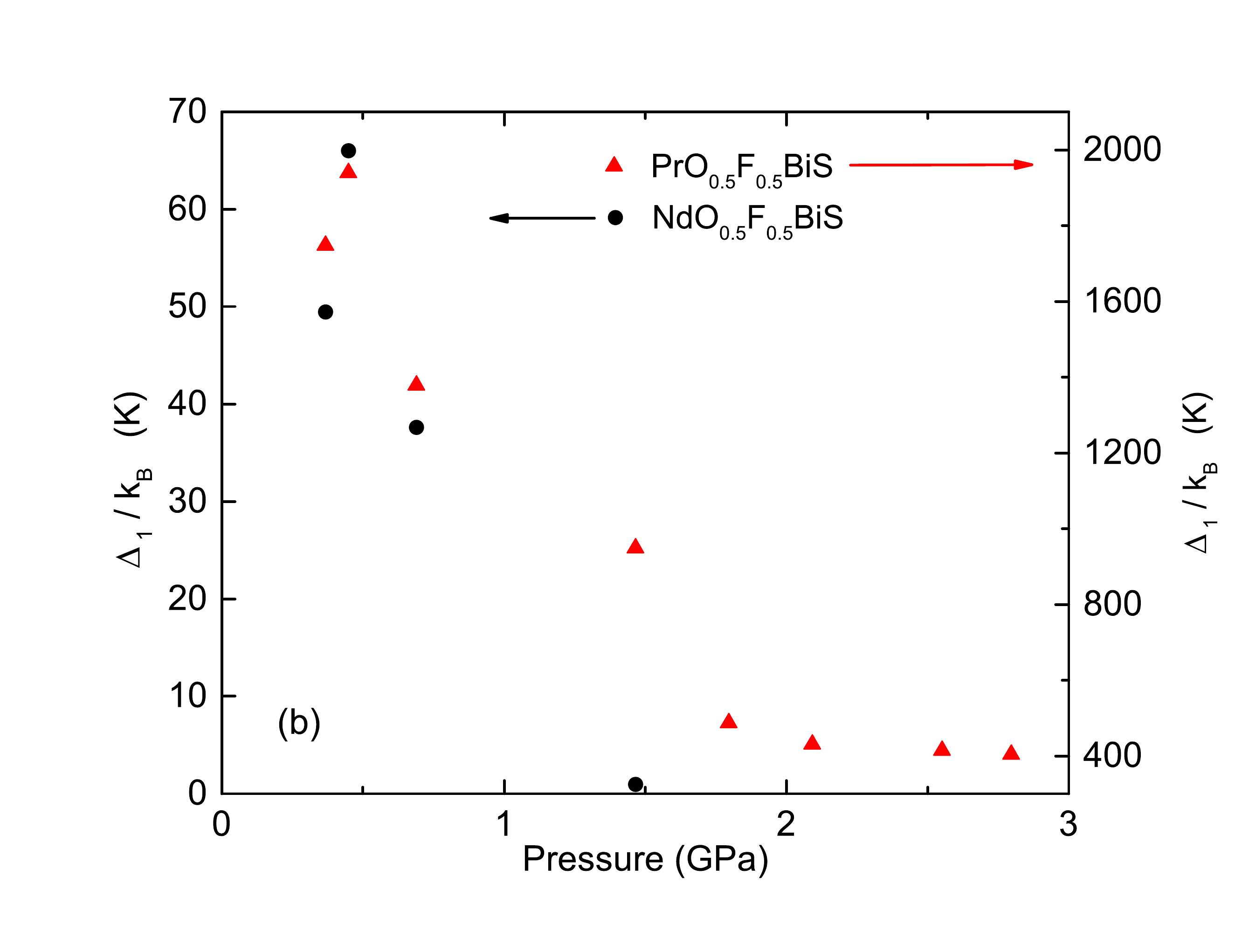}
\caption{\label{Resistivity2}  (colour online)~(a) Electrical resistivity $\rho$ vs pressure in the normal state (plotted on a $\log$ scale). Values of $\rho$ for PrO$_{0.5}$F$_{0.5}$BiS$_{2}$ and NdO$_{0.5}$F$_{0.5}$BiS$_{2}$ were taken at \textit{T} = 8.5 and 7.5 K, respectively. Dotted lines reflect the slopes (suppression rates), and arrows point to changing slopes at $\sim$2.1 GPa. The breaks in slope occur at the transition pressure $P_\mathrm{t}$. (b) Energy gap $\Delta_1$ ($\sim$100 - 200 K) vs pressure for both compounds. The rate of decrease in $\Delta_1$ with pressure flattens at $\sim$2 GPa in both compounds. Above $\sim$1.9 GPa, $\Delta_1$/k$_B$ = 0 K in NdO$_{0.5}$F$_{0.5}$BiS$_{2}$.}
\end{figure}
%\end{comment} 
In the NdO$_{0.5}$F$_{0.5}$BiS$_{2}$ compound, there is a significant jump in \textit{$T_\mathrm{c}$} from 3.9 to 6.3 K as pressure is increased from $\sim$1.8 to 2.1 GPa (d$\textit{$T_\mathrm{c}$}$/d$P$ = 8.0 K GPa$^{-1}$). Following the rapid increase in \textit{$T_\mathrm{c}$}, both compounds exhibit a high \textit{$T_\mathrm{c}$} superconducting phase, in which the evolution of \textit{$T_\mathrm{c}$} exhibits a domelike behaviour; i.e., \textit{$T_\mathrm{c}$} gradually increases to its maximum value and then slowly decreases with pressure. In the PrO$_{0.5}$F$_{0.5}$BiS$_{2}$ compound, \textit{$T_\mathrm{c}$} increases to a maximum value of 7.6 K at 2.5 GPa and then steadily decreases with pressure, while in the NdO$_{0.5}$F$_{0.5}$BiS$_{2}$ compound, \textit{$T_\mathrm{c}$} increases to a maximum \textit{$T_\mathrm{c}$} of 6.4 K at $\sim$2.5 GPa and then decreases slowly with pressure.\\ 
\indent In the normal state (above the \textit{$T_\mathrm{c}(P)$} curves shown in Figure~\ref{Tc Combined Spline}), the semiconducting behaviour in both compounds is continuously suppressed with pressure as manifested by the decrease of the energy gap $\Delta_1$  (defined below) with pressure, the values of which are indicated in the false colour legend of Figure~\ref{Tc Combined Spline}(a) and (b) for PrO$_{0.5}$F$_{0.5}$BiS$_{2}$ and NdO$_{0.5}$F$_{0.5}$BiS$_{2}$, respectively. The NdO$_{0.5}$F$_{0.5}$BiS$_{2}$ sample exhibits a fully metallic state at $\sim$2 GPa (where $\Delta_1$ vanishes), represented by the green region to the right of the crosshatching in Figure~\ref{Tc Combined Spline}(b).\\
\indent From the plot of $\log(\rho)$ vs. $P$ displayed in Figure~\ref{Resistivity2}(a), there is a noticeable change in the magnitude of the suppression rate, d$\log(\rho)$/d$P$, for both the PrO$_{0.5}$F$_{0.5}$BiS$_{2}$ and NdO$_{0.5}$F$_{0.5}$BiS$_{2}$ compounds.  The $\rho(P)$ data for PrO$_{0.5}$F$_{0.5}$BiS$_{2}$ and NdO$_{0.5}$F$_{0.5}$BiS$_{2}$ were taken in the normal state at 8.5 K and 7.5 K, respectively. These temperatures occur just above the onset of the superconducting transition at \textit{$T_c^{\mathrm{onset}}$}. In both compounds, there is a strong suppression of electrical resistivity up to $\sim$2.1 GPa, followed by a weaker suppression at higher pressures. The dotted lines in Figure~\ref{Resistivity2}(a) are guides to the eye for the rates of suppression. The change in the suppression rate near 2.1 GPa (emphasized by the vertical arrows in Figure~\ref{Resistivity2}(a)) is coincident with $P_\mathrm{t}$.\\
\indent The semiconducting behaviour of the $\rho(T)$ data and its rapid suppression with pressure was noted by Kotegawa \textit{et al}. in their study of the LaO$_{0.5}$F$_{0.5}$BiS$_{2}$ compound synthesized under high pressure.\cite{Kotegawa} They observed that $\rho(T)$ could be described over two distinct temperature regions by the relation $\rho(T)$ = $\rho_0\rme^{\Delta/2k_{B}T}$ where $\rho_0$ is a constant and $\Delta$ is an energy gap. In a recent paper, we applied this analysis to extract the high and low temperature energy gaps $\Delta_1$ and $\Delta_2$ for the \textit{Ln}O$_{0.5}$F$_{0.5}$BiS$_2$ (\textit{Ln} = La, Ce) compounds.\cite{Wolowiec} We used the same analysis in the current study to determine the value of the high temperature energy gap $\Delta_1$ for both compounds \textit{Ln}O$_{0.5}$F$_{0.5}$BiS$_2$ (\textit{Ln} = Pr, Nd). The energy gap $\Delta_1$ in NdO$_{0.5}$F$_{0.5}$BiS$_{2}$ was determined using the $\rho(T)$ data from the region 100 - 200 K for lower pressures 0.37 - 0.69 GPa and $\rho(T)$ data from the region 20 - 100 K for higher pressures 1.47 - 2.80 GPa.  For PrO$_{0.5}$F$_{0.5}$BiS$_{2}$, the energy gap $\Delta_1$ was extracted using $\rho(T)$ data in the region 200 - 280 K for all pressures up to 2.80 GPa.\\
\indent The pressure dependence of the energy gap $\Delta_1$ for both compounds is shown in Figure~\ref{Resistivity2}(b). The energy gap $\Delta_1$ decreases rapidly with pressure up to $\sim$2 GPa. Above $\sim$2 GPa, $\Delta_1$ exhibits relatively little pressure dependence. This is consistent with the transition to a weaker suppression rate shown in Figure~\ref{Resistivity2}(a) which also sets in at $\sim$2 GPa.  In NdO$_{0.5}$F$_{0.5}$BiS$_{2}$, the energy gap $\Delta_1$/$k_B$ = 0 K above 1.8 GPa. This is consistent with the semiconductor-metal transition near 2 GPa indicated by a positive temperature coefficient of electrical resistivity (d$\rho$/d$T > $ 0) seen in the $\rho(T)$ data shown in Figure~\ref{Resistivity}(b). The rapid decrease in the energy gap $\Delta_1$ for $P < P_t$ in both the \textit{Ln}O$_{0.5}$F$_{0.5}$BiS$_2$ (\textit{Ln} = Pr, Nd) compounds is similar to behaviour observed previously in the \textit{Ln}O$_{0.5}$F$_{0.5}$BiS$_2$ (\textit{Ln} = La, Ce) compounds.\cite{Kotegawa,Wolowiec}\\ 
\section{Discussion}
\label{sec:discussion}
\begin{figure}[h!]
\centering
\includegraphics[scale=0.38, trim= 2.3cm 1.2cm 0cm 2.1cm, clip=true]{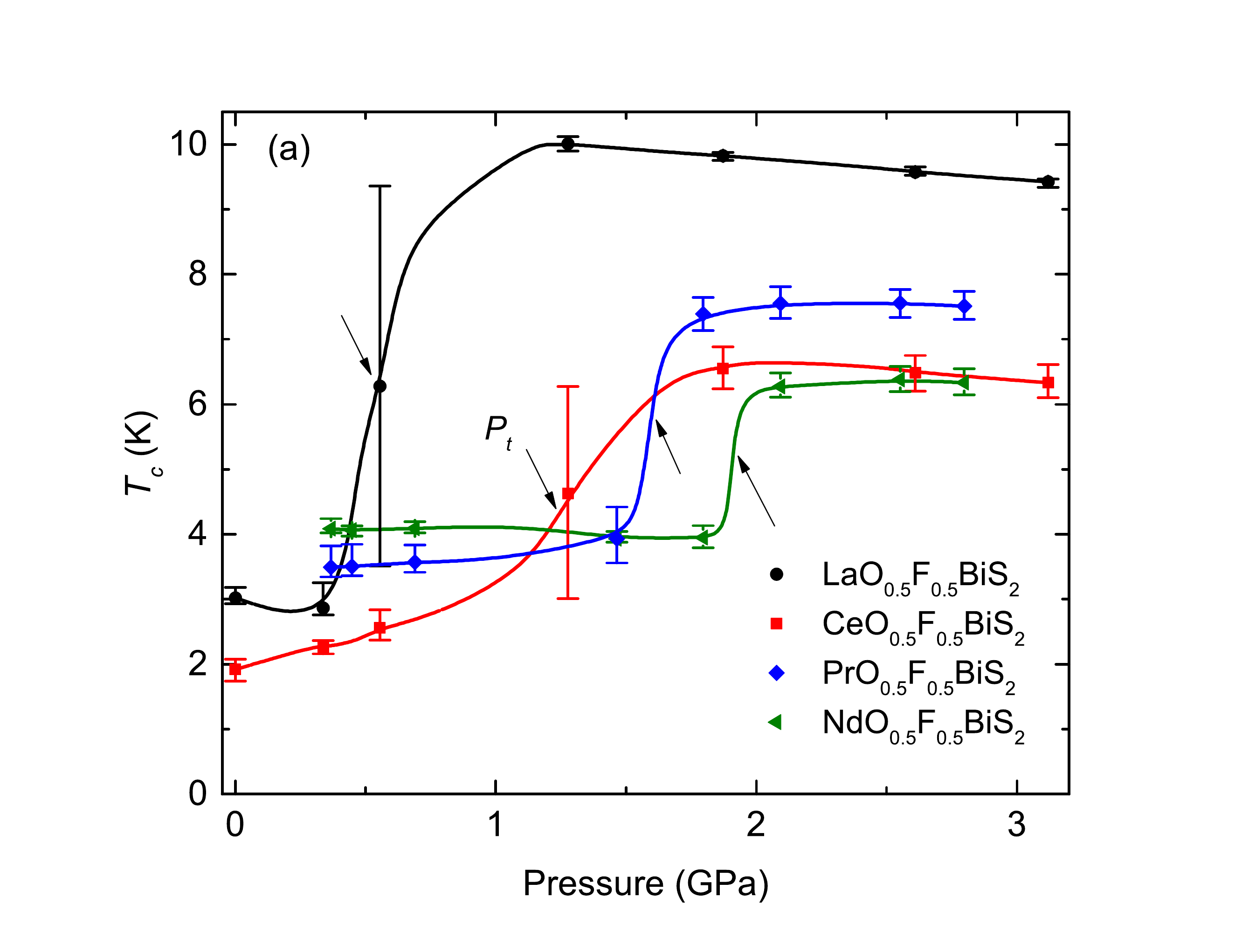}\\
\includegraphics[scale=0.38, trim= 2.3cm 1.0cm 0cm 1.4cm, clip=true]{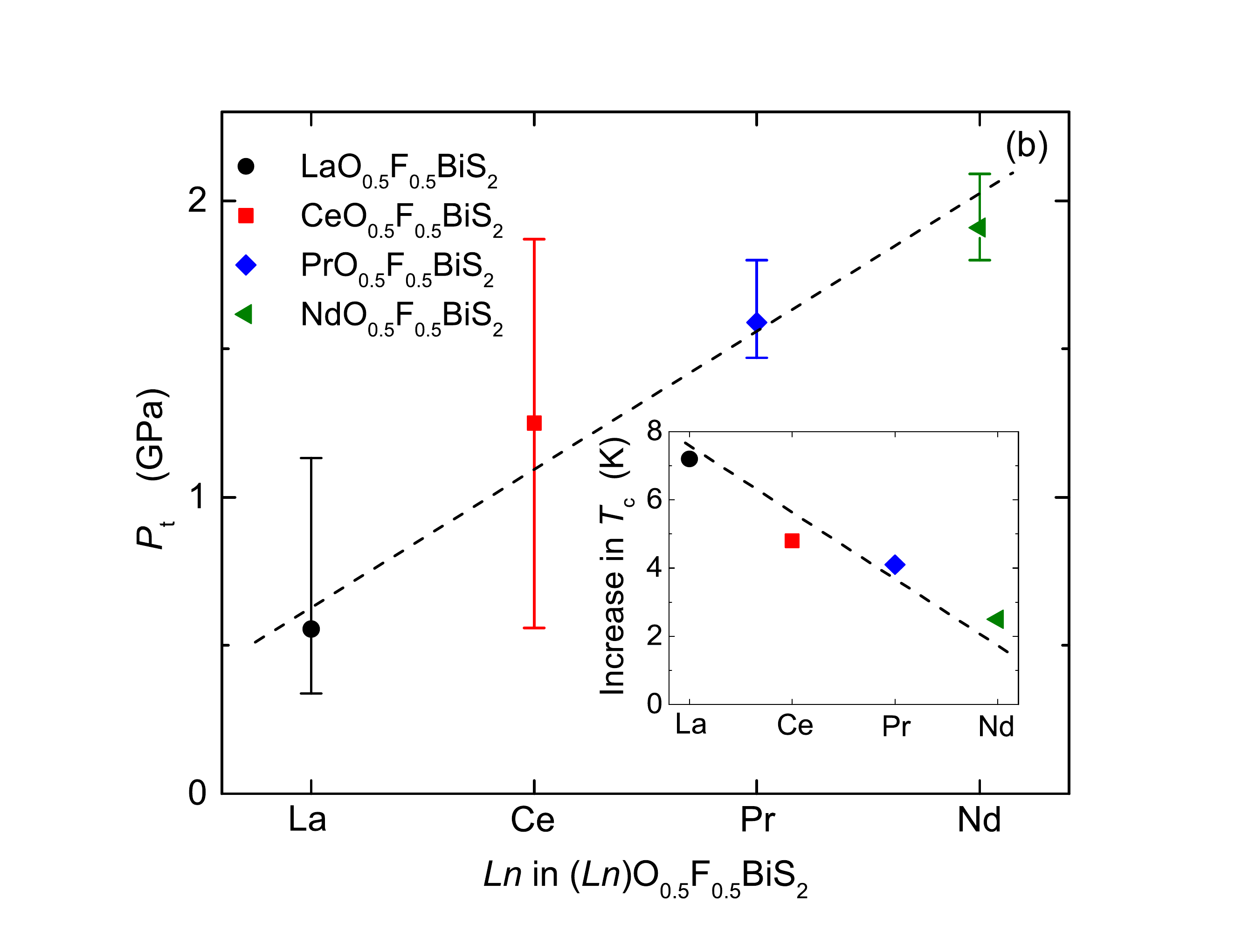}
\caption{\label{Tc_Summary}  (colour online) (a) \textit{$T_\mathrm{c}$} vs pressure plotted for the four compounds \textit{Ln}O$_{0.5}$F$_{0.5}$BiS$_{2}$ (\textit{Ln} = La, Ce, Pr, Nd). The black arrows emphasize the transition pressure $P_\mathrm{t}$ which is defined in the text. (b) Transition pressure $P_\mathrm{t}$ plotted as a function of \textit{Ln} in \textit{Ln}O$_{0.5}$F$_{0.5}$BiS$_{2}$. The inset displays the increase in \textit{$T_\mathrm{c}$} at $P_\mathrm{t}$ as a function of \textit{Ln}. Dashed lines are guides to the eye.}
\end{figure} 
\indent Both the temperature dependence of the electrical resistivity $\rho$ and the evolution of \textit{$T_\mathrm{c}$} under applied pressure for the \textit{Ln}O$_{0.5}$F$_{0.5}$BiS$_2$ (\textit{Ln} = Pr, Nd) samples reported in section~\ref{sec:results} of this paper are markedly similar to the results we recently reported for the \textit{Ln}O$_{0.5}$F$_{0.5}$BiS$_2$ (\textit{Ln} = La, Ce) compounds.\cite{Wolowiec} As shown in the phase diagrams displayed in Figure~\ref{Tc_Summary}(a), each of the four compounds \textit{Ln}O$_{0.5}$F$_{0.5}$BiS$_2$ (\textit{Ln} = La, Ce, Pr, Nd) exhibits an abrupt pressure-induced transition from a low \textit{$T_\mathrm{c}$} superconducting phase at lower pressure to a high \textit{$T_\mathrm{c}$} superconducting phase at higher pressure.\\
\indent In the four compounds, the pressure-induced transition observed in the superconducting state is coincident with changes in the suppression of the electrical resistivity $\rho$ in the normal state. The rate of suppression of semiconducting behaviour (Figure~\ref{Resistivity2}(a)) and the rate of decrease in the energy gap $\Delta_1$ (Figure~\ref{Resistivity2}(b)) both saturate at pressures that correlate with the transition pressure into the high \textit{$T_\mathrm{c}$} superconducting phase. In the specific case of the NdO$_{0.5}$F$_{0.5}$BiS$_{2}$ compound, a semiconductor-metal transition occurs at $P_\mathrm{t}$.  These changes in the normal state electrical resistivity indicate there may be significant increases in the charge carrier density during the rapid increase in \textit{$T_\mathrm{c}$} that occurs between the two superconducting phases.\cite{Igawa}\\ 
\indent The transition pressures $P_\mathrm{t}$, indicated by the black arrows in the temperature-pressure phase diagrams of Figure~\ref{Tc_Summary}(a), were defined as the pressure corresponding to the value of \textit{$T_\mathrm{c}$} at the midpoint between the values of \textit{$T_\mathrm{c}$} in the low and high \textit{$T_\mathrm{c}$} phases immediately preceding and following the transition. $P_\mathrm{t}$ is plotted as a function of lanthanide element (\textit{Ln} = La, Ce, Pr, Nd) in \textit{Ln}O$_{0.5}$F$_{0.5}$BiS$_2$ in Figure~\ref{Tc_Summary}(b). There is a clear linear relationship between the increasing atomic number of \textit{Ln} and an increase in $P_\mathrm{t}$. The magnitude of the ``jump'' in \textit{$T_\mathrm{c}$} also scales with the atomic number of the \textit{Ln} element in \textit{Ln}O$_{0.5}$F$_{0.5}$BiS$_2$ as clearly shown in the inset of Figure~\ref{Tc_Summary}(b). The pressure-induced increase in \textit{$T_\mathrm{c}$} decreases in magnitude as the atomic number of the lanthanide element increases. The lengths of the vertical bars in Figure~\ref{Tc_Summary}(b) represent the respective pressure windows over which the transitions from the low \textit{$T_\mathrm{c}$} phase to the high \textit{$T_\mathrm{c}$} phase occurred in each of the four compounds. The pressure range over which the transition occurs also decreases with increasing atomic number of \textit{Ln} in \textit{Ln}O$_{0.5}$F$_{0.5}$BiS$_2$. 
Table~\ref{Table:Transition} contains values of low \textit{$T_\mathrm{c}$}, maximum \textit{$T_\mathrm{c}$}, overall increase in \textit{$T_\mathrm{c}$} (max. \textit{$T_\mathrm{c}$} - low \textit{$T_\mathrm{c}$}), and transition pressure $P_\mathrm{t}$, for each compound.\\
\begin{table}[t]
\caption{\textit{$T_\mathrm{c}$} data for \textit{Ln}O$_{0.5}$F$_{0.5}$BiS$_2$ (\textit{Ln} = La, Ce, Pr, Nd).}          
\centering     
    \begin{tabular}{ c  c  c  p{3.3cm}  c }
     \br
     \textit{Ln} & ~low \textit{$T_\mathrm{c}$} (K) & ~~max. \textit{$T_\mathrm{c}$} (K) & increase~in~\textit{$T_\mathrm{c}$} (K) & $P_\mathrm{t}$ (GPa)\\ %[.9ex]
     \mr
    La  & 2.9   & ~10.1  & ~~~~~~~~~7.2  & 0.56  \\ 
    Ce  & 1.9  & ~~6.7  & ~~~~~~~~~4.8   &  1.25 \\ 
    Pr  & 3.5   & ~~7.6  & ~~~~~~~~~4.1    & 1.59  \\
    Nd  & 3.9  & ~~6.4 & ~~~~~~~~~2.5   & 1.91 \\ [1.3ex]
    \br
    \end{tabular}
    \label{Table:Transition}
    \end{table}
\indent The evolution of \textit{$T_\mathrm{c}$} with pressure in the low \textit{$T_\mathrm{c}$} phase has recently been reported for the \textit{Ln}O$_{0.5}$F$_{0.5}$BiS$_2$ (\textit{Ln} = Pr, Nd) compounds in two studies by Selvan \textit{et al}.\cite{Selvan,Selvan2}  For PrO$_{0.5}$F$_{0.5}$BiS$_{2}$, they observed a gradual increase of \textit{$T_\mathrm{c}$} from 3.7 to 4.7 K with pressure up to $\sim$2.2 GPa.\cite{Selvan2} For NdO$_{0.5}$F$_{0.5}$BiS$_{2}$, they found a gradual evolution of \textit{$T_\mathrm{c}$} with pressure from 4.6 to 5.0 K up to a pressure of $\sim$1.3 GPa and then down from 5.0 to 4.8 K upon further application of pressure up to $\sim$1.8 GPa.\cite{Selvan} In both reports, however, there was no evidence of a transition characterized by a rapid increase in \textit{$T_\mathrm{c}$}. It is possible that slight variations in the chemical composition of the samples in their studies compared to those studied by us may be responsible for differences in the material's response to applied pressure. Furthermore, the pressure transmitting media used in this study (see section~\ref{sec:experiment}) and the transmitting fluid used in their studies \cite{Selvan,Selvan2} may have different properties with regard to pressure gradients that can affect the measured pressure at which a transition occurs.\cite{Butch} It is also possible that the pressures reached in their studies were lower than the pressure required to induce the transitions that we observed in this report.\\  
\section{Concluding remarks}
\label{sec:concluding remarks}
\indent We have observed markedly similar behaviour in the temperature dependence of the normal state electrical resistivity and evolution of the superconducting critical temperature \textit{$T_\mathrm{c}$} under applied pressure for the two BiS$_2$-based superconductors \textit{Ln}O$_{0.5}$F$_{0.5}$BiS$_2$ (\textit{Ln} = Pr, Nd).  The qualitative behaviour observed for the two compounds in this study is strikingly similar to the results we recently reported for the two BiS$_2$-based superconductors \textit{Ln}O$_{0.5}$F$_{0.5}$BiS$_2$ (\textit{Ln} = La, Ce).\cite{Wolowiec} In each of the four compounds \textit{Ln}O$_{0.5}$F$_{0.5}$BiS$_2$ (\textit{Ln} = La, Ce, Pr, Nd), there is a sizable enhancement of \textit{$T_\mathrm{c}$} in the superconducting state accompanying the suppression of semiconducting behaviour with pressure in the normal state. The suppression of the semiconducting behaviour in the normal state saturates at a critical pressure $P_\mathrm{t}$ which corresponds to the pressure where there is a transition between a  low \textit{$T_\mathrm{c}$} superconducting phase and a high \textit{$T_\mathrm{c}$} superconducting phase. The semiconducting behaviour of the electrical resistivity in the normal state is consistent with an energy gap that is suppressed with pressure in a similar way.  In the particular case of the NdO$_{0.5}$F$_{0.5}$BiS$_2$ compound, there is a pressure-induced semiconductor-metal transition at $P_\mathrm{t}$ $\approx$ 2 GPa.\\
\indent The coincidence of the saturation of the suppression of semiconducting behaviour in the normal state electrical resistivity with the rapid increase in \textit{$T_\mathrm{c}$} indicates there may be significant increases in the charge carrier density in the vicinity of the pressure-induced transition. We found that the transition pressure $P_\mathrm{t}$ (see section~\ref{sec:discussion}) increases with increasing atomic number of the lanthanide element (\textit{Ln} = La, Ce, Pr, Nd) in \textit{Ln}O$_{0.5}$F$_{0.5}$BiS$_2$. However, the size of the increase in \textit{$T_\mathrm{c}$} between the two superconducting phases decreases as lanthanide atomic number increases. The scaling of both the transition pressure $P_\mathrm{t}$ and the size of the ``jump'' in \textit{$T_\mathrm{c}$} with the atomic number of the lanthanide element suggests that the pressure-induced transition between the two superconducting phases may be a structural transition; however, at present, the precise mechanism driving the enhancement of \textit{$T_\mathrm{c}$} with pressure is unknown. X-ray diffraction experiments under pressure on the LaO$_{0.5}$F$_{0.5}$BiS$_{2}$ compound are currently underway to help determine whether the pressure-induced enhancement of \textit{$T_\mathrm{c}$} and the suppression of semiconducting behaviour are related to a structural transition.\\ 

\ack
The authors thank A. J. Friedman for his helpful work in the synthesis of the \textit{Ln}O$_{0.5}$F$_{0.5}$BiS$_2$ (\textit{Ln} = La, Ce, Pr, Nd) compounds. High pressure research at the University of California, San Diego (UCSD) was supported by the National Nuclear Security Administration under the Stewardship Science Academic Alliance Program through the U.S. Department of Energy (DOE) under Grant No. DE-NA0001841.  Sample synthesis at UCSD was sponsored by the U.S. Air Force Office of Scientific Research under MURI Grant No. FA9550-09-1-0603. Characterization of samples at ambient pressure was supported by the U.S. DOE Grant No. DE-FG02-04-ER46105.

\end{document}